\documentclass[twoside]{LCWS11}
\usepackage[latin1]{inputenc}
\usepackage[dvips]{graphicx,epsfig,color}
\usepackage{wrapfig,rotating}
\usepackage{amssymb,amsmath,array}
\begin{document}
\title{
Report on the Forward Tracking Detector of the International Large Detector} 
\author{Alberto Ruiz-Jimeno \\
(on behalf of the Forward Tracking Detector for ILD Group)
\vspace{.3cm}\\
\thanks{The reported work was funded under MICINN projects CSD2007-0042
and FPA2008-03564-E/FPA.}
\vspace{.3cm}\\
IFCA, Instituto de F\'isica de Cantabria ( CSIC-Universidad de Cantabria)\\
Avda. los Castros, s/n
39005 Santander- Spain
}

\maketitle

\begin{abstract}
An overview of some progress and R\&D activities of the Spanish network for future accelerators concerning the forward tracker detector of the International Large Detector is shown.
\end{abstract}

\section{Introduction}

\begin{figure}[h]
\centerline{\includegraphics[width=1.\columnwidth]{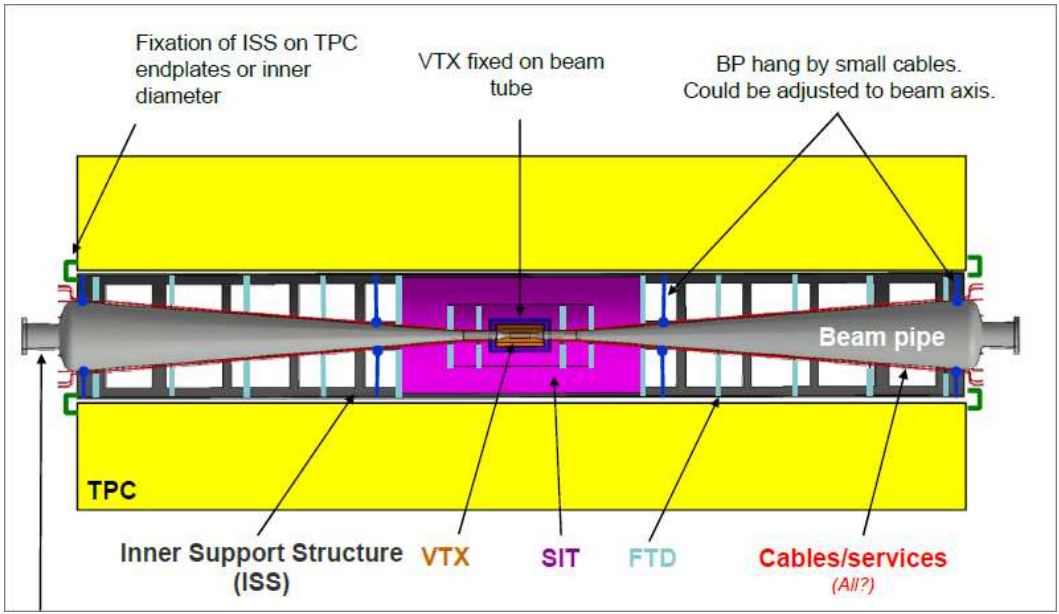}}
\caption{Integration of the inner detector of ILD } \label{Fig:Integration}
\end{figure}

 The Spanish groups~\cite{Spanish_network} are developing design studies for the Forward Tracker Detector (FTD) of the International Large Detector (ILD) concept both on sensor technologies, mechanics, readout electronics, power supply, alignment, integration and monitoring. The integration of the Inner Detector (Figure~\ref{Fig:Integration}) is a critical item to achieve accurate and efficient reconstruction of charged particle trajectories as well as primary and secondary vertexes.This report gives a glimpse of the advances performed on integration and software modeling as well as R\&D working lines.

\section{FTD progress on Integration and Software modeling}
\begin{figure}[h]
\vspace{-2cm}
\centerline{\includegraphics[width=1.\columnwidth]{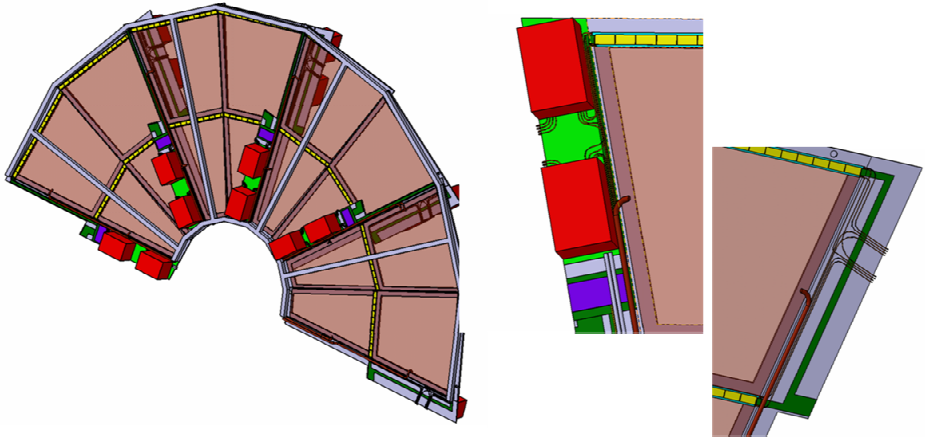}}
\vspace{-1cm}
\caption{Portion of a disk mechanical design and detail of power distribution system, readout, and cables}\label{Fig:Disk}
\end{figure}

A detailed study of realistic mechanics, including power distribution system, r/o electronics, cables and optical links (not final design concerning material budget) has been performed (Figure~\ref{Fig:Disk}).
A software modeling driver~\cite{Jordi} of the FTD has been included in the simulation program (MOKKA)~\cite{MOKKA} and a new interface for the Geometry reconstruction package GEAR~\cite{GEAR} has been implemented. This interface includes persistency for digitization and reconstruction.

The Mokka driver includes:
\begin {itemize}
    \item Self-scaling: the significant parameters and positioning are defined versus the surrounding components of the inner subcomponents vertex detector (VTX), Silicon Barrel internal tracker (SIT), Silicon Barrel external tracker (SET), beam pipe,..., following the Letter of Intent~\cite{LOI} specifications.
    \item The sensitive areas are placed over petals generating disk structures. The disk 1 and 2 are composed of pixels and the disk 3 to 7 are composed of microstrips; in this case the read-out chips and hybrid structure is included.
    \item Two designs are available. The "per-default" is an staggered design. Another option is a turbine-like design.
    \item The carbon-fiber inner cylinder supports the whole disks structure w.r.t. the beam tube. The outer cylinder supports the 4-7 microstrip disks w.r.t the Time  projection Chamber (TPC).
    \item The cables are located in the inner cylinder as a cone. It needs to be updated.
\end{itemize}

The FTD digitization has the geometrical interface decoupled, so every microstrip based detector could use potentially the code. The code is currently used for the digitization of the vertex subdetector (SVD) in Belle II~\cite{BelleII} and incorporates digitization of microstrips and cluster finding algorithms based on centre of gravity or on head-tail analog methods. The digitization includes drift in the electric field, diffusion due to multiple collisions, Lorentz shift in magnetic field, mutual microstrip crosstalks and electronic noise. The clustering transforms electric pulses into real hits. The code is integrated in the reconstruction framework Marlin~\cite{Marlin}. The code for FTD is currently at a debugging phase.
\\
\section{R \& D lines of work}

 Several R\&D activities suitable for FTD has been done on microstrip sensors and monitoring.

 Silicon tracking sensors with minor modifications to make them highly transparent to infrared light will be used for internal alignment. We have developed a realistic simulation of the microstrip structures taking into account both interferential and diffractional effects. We have used a Rigorous Coupled Wave Analysis (RCWA) to solve Maxwell equations exactly and predict the transmitted electromagnetic fields~\cite{RCWA}. We then applied a Fresnel approximation to calculate the far-field, several centimeters away from the grating. It was the first time that an optical simulation of these sensors has included this level of detail. The simulation was validated comparing its result with known diffraction samples. We obtained an increase of +30\% in transmittance with respect to non-optimized sensors which means an absolute transmittance of 50\%~\cite{Transparency, NIM}.

 We are performing studies for structural and environmental monitoring of tracker and vertex systems using Fiber Optic Sensors (FOS)~\cite{Moya}. The motivation is to achieve real time monitoring of environment variables as temperature, humidity, CO2, magnetic field, as well as real time structural monitoring of deformations, vibrations (push \& pull operation) and any movement by using technologies with low material budget and high multiplexing capability. The FOS incorporate Fiber Bragg Grating optical transducers, are very light and have immunity against high electromagnetic fields, high voltages and high and low temperatures. They can be embedded in material composite and the signal of the sensor is encoded in the wavelength, making measurements transferable and neutrals to intensity drifts. The sensors are manufactured in different fibers type (depending on manufacturing procedure) mainly with three different coatings, Acrylate, Polyimide and Ormocer. As nuclear radiation environments can induce changes of the mechanical properties changing the sensor grating period, calibration with the different fibers type and coatings need to be explored. The FOS are flexible, have a reasonable cost and have multiplexing capability and low loss and long-range signal transmission. We are analyzing different coatings, as well as small samples of composite laminates with different stack configurations.  We have also manufactured and calibrated a  mechanical mock-up to monitor the temperature and position for the PXD subdetector of Belle II~\cite{BelleII}. Test of irradiated samples are also being performed.

 Another activity concerns a novel 2D position sensitive semiconductor detector concept~\cite{Francisca}, based on simple single-sided AC-coupled
microstrip detectors with resistive coupling electrodes. The feasibility method has been demonstrated with different prototypes tested with a Near Infrared laser and validated against a simulation of sensor's equivalent circuit. We have achieved excellent agreement between experimental and simulation data.
Prototypes have been also tested at SPS pion beam at CERN. We are analyzing the results and new test beams and laser characterization are in
progress.

\section{Bibliography}

\begin{footnotesize}

\end{footnotesize}

\end{document}